\documentclass[twocolumn,prl]{revtex4}
\usepackage{amsmath}
\usepackage{amssymb}
\usepackage{graphicx}
\usepackage{psfrag}
\usepackage{subfigure}

\def\be{\begin{equation}}
\def\ee{\end{equation}}
\def\bem{\begin{pmatrix}}
\def\eem{\end{pmatrix}}
\def\del{\partial}
\def\ol{\overline}
\def\ed{\epsilon_{\rm d}}
\def\eq#1{Eq.~(\ref{#1})}
\def\fig#1{Fig.~\ref{#1}}
\def\Lag{\mathcal{L}[\sigma]}
\DeclareMathOperator{\Tr}{Tr}
\DeclareMathOperator{\tr}{tr}
\DeclareMathOperator{\Trg}{Trg}
\DeclareMathOperator{\trg}{trg}
\DeclareMathOperator{\diag}{diag}
\DeclareMathOperator{\e}{e}
\DeclareMathOperator{\dif}{d}

\begin{document}

\title{Persistent current on a disordered mesoscopic ring with an embedded quantum dot}
\author{Erlon Henrique {Martins Ferreira}}\email{erlon@fisica.ufmg.br} 
\author{Maria Carolina {Nemes}}
\affiliation{Departamento de F{\'\i}sica, ICEx, UFMG, \\
P.O.Box 702, 30.161-970, Belo Horizonte, MG, Brazil}
\author{Hans A. {Weidenm{\"u}ller}}
\affiliation{Max Planck Institut f{\"u}r Kernphysik, \\
D-69029, Heidelberg, Germany}
\date{\today}
\begin{abstract}
We calculate the ensemble averaged persistent current on disordered mesoscopic rings with an embedded quantum dot. We model the quantum dot as a single resonance and use Random Matrix Theory to model the impurities in the ring. Using Efetov's supersymmetry technique, we develop an analytical expression for the current. We find not only one, but two resonance peaks in the current. This is interpreted as quantum interference phenomenon.
\end{abstract}
\maketitle
\section{Introduction}

Given the rapid development of the technology in the fabrication and manipulation of semiconductor structures, there has been 
great interest in the study of different mesoscopic systems, specially involving quantum dots. Many atomic effects as the Kondo and Fano effects can also be observed in these systems and in controllable way\cite{alhassid}. Those work deal mainly with the well known Aharonov-Bohm interferometer, which is basically a ring attached to at least two external leads. Then it is possible to measure, for instance, the interference pattern of the electronic wave function as a function of an external magnetic field, or the dephasing of this very function by the tunneling of the electron through the quantum dot, among many other features.

We are here basically interested in an isolated ring with an embedded quantum dot, and its effect on the persistent current. Some recent works\cite{xiong1,xiong2,kang,ferrari,kobayashi, hewson, ng,glazman} have driven attention to this problem, specially in the Kondo regime. However none of them consider the disorder in the ring, and are solved numerically by means of the slave-boson mean-field theory. We propose a new approach to the problem, and also find some interesting typically quantum features of the current that were not observed before.

\section{Supersymmetric approach}

\begin{figure}
  \psfrag{q}[][]{QD}
  \psfrag{e}[][]{$\ed$}
  \psfrag{v}[][]{\ $v$}
  \psfrag{f}[][]{$\Phi$}
\includegraphics{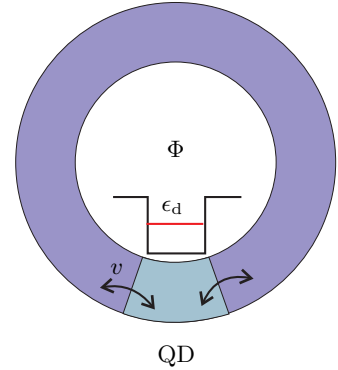}
\caption{Aharonov-Bohm ring with an embedded quantum dot.}
\label{fig1}
\end{figure}

Following the IWZ model\cite{iwz}, we consider the simplest case and divide our system into two boxes: one of the quantum dot (QD) and the other of the ring (see \fig{fig1}). Only the ring has a disorder given by a random GOE Hamiltonian $H_\text{GOE}$ of rank $N$ and second momentum given by
\be
\overline{H_{\mu \nu} H_{\mu' \nu'}} = \frac{\lambda^2}{N} [\delta_{\mu \mu'} \delta_{\nu \nu'} + \delta_{\mu\nu'} \delta_{\nu \mu'} ]\;.
\ee
$2\lambda$ is the radius of the Wigner semicircle. The QD is assumed to have one single level $\ed$. We include the Aharonov-Bohm (AB) phase in the coupling term between the walls of the QD and the ring. Doing so, the coupling is given by
\be v(\phi)=v_L\e^{i\phi}+v_R\e^{-i\phi} = v_0\cos(\phi)\,.\ee
Here $v_{L,R}$ are the coupling through the left and right side of the dot, which we can consider essentially the same. The AB phase is $\phi=\pi\Phi/\Phi_0$, with $\Phi_0=hc/e$. In this way, the Hamiltonian of the system has dimension  $(1+N)\times(1+N)$ and is given by
\be H = \begin{pmatrix} \epsilon & V \\ V^T & H_\text{GOE} \end{pmatrix} \ee\ .
$V$ is a $N$ dimensional vector with constant elements $v(\phi)$.

The persistent current in the ring can be written in terms of the eigenenergies $E_n$ of the system as 
\be I(\phi) = -\frac{2\pi c}{\Phi_0} \sum_{n}\frac{\del E_n}{\del \phi}\;, \ee
with the condition that $\max{E_n}<E_F$, $E_F$ being the Fermi energy. It is more convenient though to write this expression using Green functions. In doing so we find
\be\begin{split}
I(\phi) & = -\frac{2\pi c}{\Phi_0}\int_0^{E_F}\dif E' \sum_{n=1}^\infty \frac{\del E_n}{\del\phi}\delta(E'-\varepsilon_n) \\
&  =  -\frac{ic}{\Phi_0}\int_0^{E_F}\Tr\biggl\{\frac{\del H}{\del\phi}[(E'^{+}-H)^{-1} - (E'^{-}-H)^{-1}]\biggr\}
\end{split}\ee
$E'^{+}$ and $E'^{-}$ carry respectively a positive and a negative infinitesimal imaginary part and the limit is implicit.

By performing the average over the impurities, we can not keep $E_F$ fixed, since it changes both with impurity realization and the magnetic flux. The correct way to overcome this difficulty was proposed in Ref.\cite{altland}, and what we have to do is make an average over the number of electrons too. This approach is ideal for treating an ensemble of rings, since due to uncertainties in their shapes, it is clear that the number of electrons is not the same in every ring, but must fluctuate about an average value $\bar N$ with an amplitude $\delta N$. This is equivalent to say that the Fermi energy is allowed to vary in an interval $[E_1,E_2]$ of length $S$ such that $S=d\delta N$, with $d$ the mean level space at the Fermi surface. Thus we have the following expression for the current:
\begin{widetext}
\be\begin{split}
\ol{I}(\phi) & = -\frac{c}{2\pi\Phi_0\delta N} \int_S\dif E \int_0^E\dif E'
 \overline{\biggl[\tr\biggl\{\frac{\del H}{\del\phi}(E'^{+}-H)^{-1}\biggr\}\tr\{(E^{-}-H)^{-1}\}+\text{c.c.}\biggl]}\;.
\end{split}\ee
\end{widetext}

The bar denotes the average over the impurities. We see here that the current is expressed as a product of an advanced and a retarded Green function, which is essentially a two-point function.

It is useful to define a generating functional $Z$ from which we can obtain the current as simple derivatives, and furthermore, we can write our Hamiltonian in a Gaussian form such the procedure of averaging can be easily performed. We follow here the standard steps in the way presented in Ref.~\cite{vwz}. This generating function must then have the form
\be
Z[j,\delta\phi] = \int\dif[\Psi]\exp\Bigl\{\tfrac i2 \Psi^\dag L^{1/2}DL^{1/2}\Psi\Bigr\}
\ee
where $\Psi$ is a graded vector of dimension $8(N+1)$ and has the form:
\be
\Psi_{\mu}^T = \bigl(S_{1\mu},S_{1\mu}^*,\chi_{1\mu},\chi_{1\mu}^*,S_{2\mu},S_{2\mu}^*,\chi_{2\mu},\chi_{2\mu}^*\bigr)\;,
\ee 
and with the definitions
\begin{align*}
D & = (E - \frac\omega 2 L -jk_2)\otimes{\bf 1}_{N+1}- \bem \epsilon\otimes{\bf 1}_8 & V \\ V^T & H_{GOE}\otimes {\bf 1}_8 \eem\, \\
V & = v_0\cos(\phi+k_1\delta\phi)\otimes{\bf e}_{1\times N}\,.
\end{align*}
Here $E=(E'^{+}+E^{-})/2$, $\omega=E^{-}-E'^{+}$ and $j$ and $\delta\phi$ are source terms. $L$, $k_1$ and $k_2$ are $8\times 8$ diagonal matrix and given by $L=\diag({\bf 1}_4,-{\bf 1}_4)$, $k_1=({\bf 1}_2,-{\bf 1}_2,0,0)$,  $k_1=(0,0,{\bf 1}_2,-{\bf 1}_2)$. Finally ${\bf 1}_{m}$ represents the identity matrix of dimension $m\times m$ and ${\bf e}_{1\times N}$ is a constant matrix of dimension  $1\times N$ and all elements equal 1.

Hence the averaged persistent current is written in terms of the functional as
\be
\ol{I}(\phi) = - \frac{cd}{8\pi S\Phi_0}\int_S\dif E \int_0^{E} \dif E' \biggl(\frac{\del^2 \ol Z}{\del\delta\phi\del j}\biggr\rvert_{\delta \phi = 0 = j} + \text{c.c.} \biggr) \ .
\ee

We may the proceed and perform the ensemble average in a standard fashion followed by the Hubbard--Stratonovitch (HB) transformation. This yields for the generating functional
\be \ol Z = \int\dif[\sigma]\exp\{\Lag\}\;,\ee
where the Lagrangian is
\be
\Lag = -\frac N4\trg\sigma^2 - \frac12\Trg\ln \ol D
\ee
and 
\be
\ol D =\bem E - \tfrac\omega2L-\epsilon-jk_2 & - V \\ -V^T & E-\tfrac\omega2L-\lambda\sigma-jk_2 \eem \;.
\ee
$\sigma$ is a $8\times 8$ graded matrix which appears as an auxiliary field to do the HB transformation. $\Trg$ denotes the graded trace, both over the $8\times8$ graded space and also the $N+1$ degrees of freedom of the system. It is understood here that the right bottom block of the above matrix has dimension $8N\times8N$.

Since we are mostly interested in the limit $N\rightarrow\infty$, we do the saddle point approximation to determine the main contribution for the exponent with $\omega=j=\delta\phi=0$. It is useful to rewrite the logarithm of $\ol D$ as 
\be
\ln\ol D  = \ln\bem \Lambda & 0 \\ 0 & \Sigma_N \eem + \ln\bem 1 & \Lambda^{-1}V \\ \Sigma_N^{-1}V^T & 1 \eem \;,
\label{lnD}
\ee
where $\Lambda = E-\epsilon-\tfrac\omega2L-jk_2$ and $\Sigma_N = (E-\lambda\sigma-\tfrac\omega2L-jk_2)\otimes{\bf 1}_N$. Expanding the second term of \eq{lnD} and taking its trace, we can rewrite it as
\be\begin{split}
\Trg\ln\bem 1 & \Lambda^{-1}V \\ \Sigma_N^{-1}V^T & 1 \eem & = 
\trg\ln\bigl[1-\Lambda^{-1}V \Sigma_N^{-1}V^T\bigr]
\end{split}\ee
Now $\trg$ runs only over the graded space.

We obtain for the saddle point equation
\be \sigma = \frac{\lambda}{E-\lambda\sigma}\left\{1+\frac1N\frac{Nv_0^2\cos^2(\phi)}{(E-\epsilon)(E-\lambda\sigma)-Nv_0^2\cos^2(\phi)}\right\}\;.\label{513} \ee
We note here that $Nv_0^2$ is of order of unity, and thus the second term in r.h.s. of \eq{513} may be dropped in the limit $N\rightarrow\infty$. We find then the diagonal solution for the saddle point equation given by
\be \sigma_{\rm D} = \frac E{2\lambda} - i\Delta L\,, \ee
where $\Delta = \sqrt{1-(E/2\lambda)^2}$. This is not however the only solution. There is actually a manifold of solutions, which can be parametrized by some group generators $T$ (see Ref.~\cite{altland2} to see its actual form), and the general solution is
\be  \sigma_{\rm sp} =  T^{-1}\sigma_{\rm D} T \equiv \frac E{2\lambda}-i\Delta Q\;. \label{ssp} \ee

We proceed with separation of $\sigma$ into the Goldstone modes $\sigma_{\rm G}$ and the massive modes $\delta\sigma$, writing $\sigma = \sigma_{\rm G} + N^{-1/2}\delta\sigma$ and expanding the logarithm in function of $\delta\sigma$. We first note that $\omega,v_0^2\sim N^{-1}$, and then make the shift $\lambda\sigma_{\rm G} \rightarrow \lambda\sigma_{\rm G} - \frac\omega2L - jk_2$, to find the effective Lagrangian
\begin{widetext}
\be\begin{split}
\mathcal{L}_{\rm ef}[\sigma]  & = -\frac{N}{4}\trg[(1-\sigma_G^2)\delta\sigma^2] + \frac{N\omega}{4\lambda}\trg(\sigma_{\rm G}L) + \frac{Nj}{2\lambda}\trg(\sigma_G k_2)
-\frac{1}{2}\trg\ln\biggl[1-\frac{Nv_0^2}{\lambda(E-\epsilon)}\cos(\phi+k_1\delta\phi)\sigma_G\cos(\phi+k_1\delta\phi)\biggr]\;.
\end{split}\ee
\end{widetext}

We have kept in the expansion only terms in first order in $\omega$ and $j$ and up to second order in $\delta\sigma$. The linear terms in $\delta\sigma$ coming from the logarithm cancel against those from $\sigma^2$. The integral over the massive modes is Gaussian and straightforward. Taking then the derivatives of the generating functional with respect to the sources $j$ e $\delta\phi$ yields
\begin{multline}
\frac{\del^2\ol{Z}}{\del j\del\delta\phi}\Bigr\rvert_{j=0=\delta\phi} =  -\frac{N^2v_0^2}{8\lambda^2}\sin(2\phi)\int\dif[\sigma_G] \exp\{\mathcal{L}_0 \}\\
\times 
\trg[\sigma_G k_2]\,\trg\biggl\{\frac{k_1\sigma_G+\sigma_Gk_1}{(E-\epsilon)-(Nv_0^2\cos^2(\phi)/\lambda)\;\sigma_G}\biggr\}\;.\label{517}
\end{multline}
where $\mathcal{L}_0 = (N\omega/4\lambda)\trg[\sigma_G L]$.

Recalling \eq{ssp}, the second trace of \eq{517} can be put into a more convenient manner as
\begin{multline}
\trg\biggl\{\frac{k_1\sigma_G+\sigma_Gk_1}{E-\epsilon-\Gamma(\phi)\,\sigma_G}\biggr\} = \\
-\frac{2\Delta^2(E-\epsilon)\trg(Qk_1)}{[E-\epsilon-(E/2\lambda)\Gamma(\phi)]^2+(\Gamma(\phi)\Delta)^2}\;.
\end{multline}
Here, $\Gamma(\phi) = Nv_0^2\cos^2(\phi)/\lambda$. The averaged persistent current is then
\begin{multline}
I(\phi) = -\frac{cd}{4\pi\Phi_0(E_2-E_1)}\frac{N^2v_0^2\Delta^2}{4\lambda^2}\sin(2\phi)\int\dif\omega \Re[f(\omega)] \\
\times \int\dif E \frac{(E-\epsilon)}{[E-\epsilon-(E/2\lambda)\Gamma(\phi)]^2+(\Gamma(\phi)\Delta)^2}\;,
\end{multline}
where
\be
f(\omega) = \int\dif[Q]\exp\Bigl\{-i\frac{N\Delta\omega}{4\lambda}\trg(QL)\Bigr\}\trg(Qk_1)\trg(Qk_2)\;.
\ee

The change of variables that takes $E^-$ and $E'^+$ into $E$ and $\omega$ implies a change in the region of integration, such that the new limits of integration are
\be
\int_0^{E_1}\dif\omega\int_{E_1-\omega/2}^{E_2-\omega/2}\dif E + \int_{E_1}^{E_2}\dif\omega\int_{\omega/2}^{E_2-\omega/2}\dif E\ .\label{519} \ee

Our model where we have used RMT to describe the disorder in the ring is valid for energies away from the boundaries of Wigner's semicircle. This implies that  $E \ll \lambda$ and therefore, we can disregard the term $E/2\lambda$ and take $\Delta=1$ in the integral over $E$. As a result, this integral becomes quite simple, and is equal to
\begin{multline}
\int\dif E \frac{(E-\epsilon)}{[E-\epsilon-(E/2\lambda)\Gamma(\phi)]^2+(\Gamma(\phi)\Delta)^2} =\\
 \frac12\ln\bigl[(E-\epsilon)^2+\Gamma(\phi)^2\bigr]\,.
\end{multline}

\section{Supersymmetric integration}

The integral $f(\omega)$ is done exactly with the parametrization for $Q$ given in Ref.~\cite{altland2} and following the same steps. This is a lengthy and monotonous task, where we have to pay special attention to the anticomuttative variables. At the end of the day, we come to the final result for the real part of $f(\omega)$
\be
\Re[f(\omega)] = 16\frac{\cos(2\pi\omega/d)-1}{(\pi\omega/d)^2}-16\;.
\ee
We have used that the mean level spacing is given by $d=\pi\lambda/N$. Defining $\Gamma_0=Nv_0^2/\lambda$, such that $\Gamma(\phi)=\Gamma_0\cos^2(\phi)$, we get the final expression for the current
 \be I(\phi) = \frac{c}{\Phi_0} \frac{\Gamma_0}{2(E_2-E_1)}
\sin(2\phi)\bigl(I_1 + I_2\bigr) \;, \label{529} \ee 
where
\begin{align*}
I_1&=\int_0^{E_1}\dif\omega\left[1+\frac{1-\cos(2\pi\omega/d)}{(\pi\omega/d)^2}\right]\\
&\qquad \times\ln\left[\frac{(E_2-\epsilon-\omega/2)^2-\Gamma(\phi)^2}{(E_1-\epsilon-\omega/2)^2-\Gamma(\phi)^2}\right]\ , \\
I_2&=\int_{E_1}^{E_2}\dif\omega\left[1+\frac{1-\cos(2\pi\omega/d)}{(\pi\omega/d)^2}\right]\\
&\qquad \times \ln\left[\frac{(E_2-\epsilon-\omega/2)^2-\Gamma(\phi)^2}{(\omega/2-\epsilon)^2-\Gamma(\phi)^2}\right]
\ .
\end{align*}

\section{Results}

\begin{figure}
\begin{picture}(0,0)%
\includegraphics[scale=0.7]{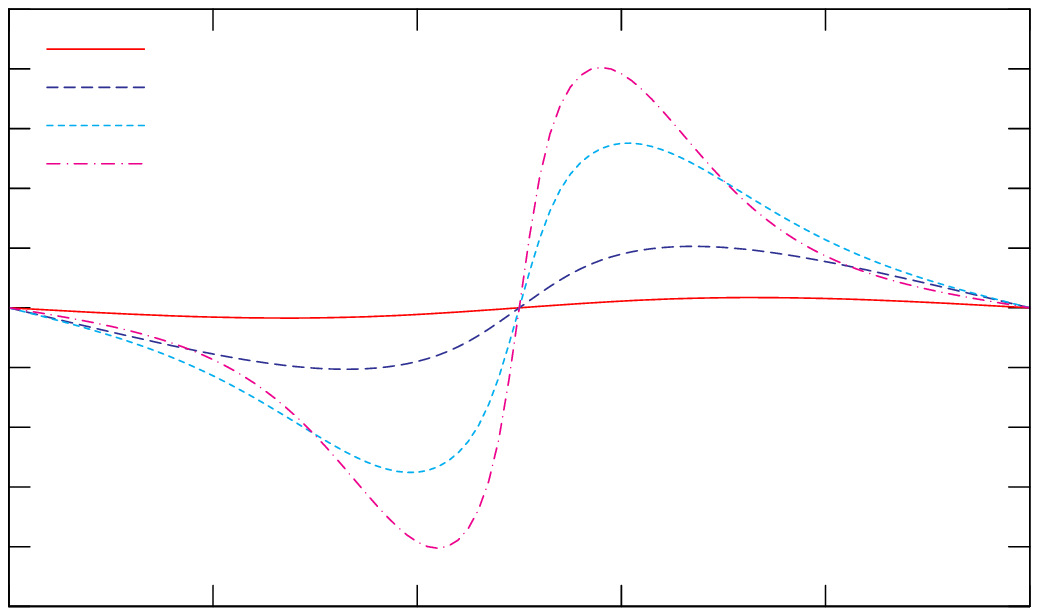}%
\end{picture}%
\begingroup
\setlength{\unitlength}{0.0140bp}%
\begin{picture}(18000,10800)(0,0)%
\put(10500,1900){\includegraphics[scale=0.25]{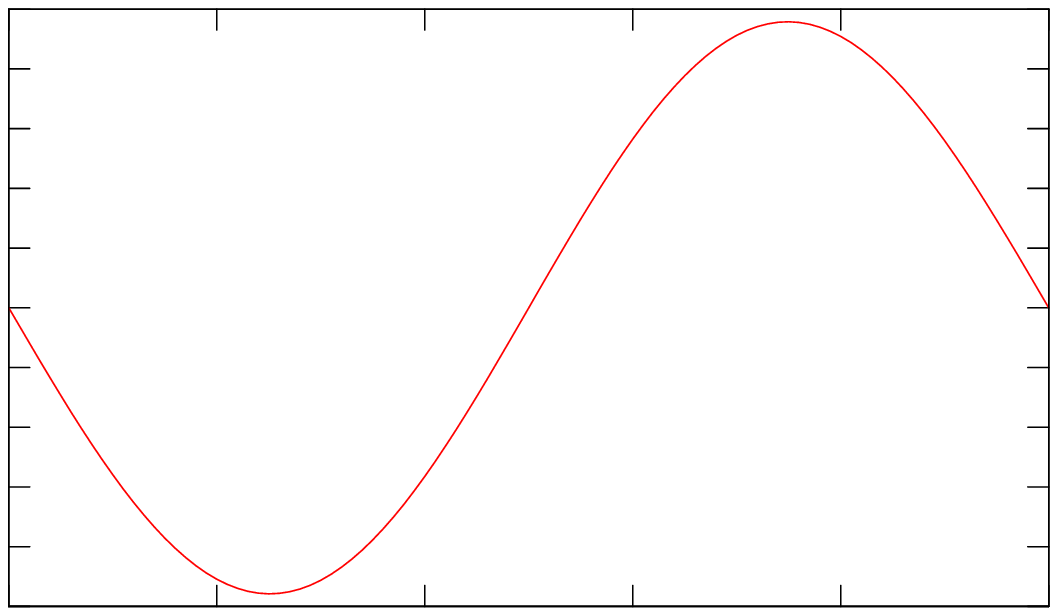}}%
\put(2200,1650){\makebox(0,0)[r]{\strut{}-2,5}}%
\put(2200,2510){\makebox(0,0)[r]{\strut{}-2,0}}%
\put(2200,3370){\makebox(0,0)[r]{\strut{}-1,5}}%
\put(2200,4230){\makebox(0,0)[r]{\strut{}-1,0}}%
\put(2200,5090){\makebox(0,0)[r]{\strut{}-0,5}}%
\put(2200,5950){\makebox(0,0)[r]{\strut{} 0}}%
\put(2200,6810){\makebox(0,0)[r]{\strut{} 0,5}}%
\put(2200,7670){\makebox(0,0)[r]{\strut{} 1,0}}%
\put(2200,8530){\makebox(0,0)[r]{\strut{} 1,5}}%
\put(2200,9390){\makebox(0,0)[r]{\strut{} 2,0}}%
\put(2200,10250){\makebox(0,0)[r]{\strut{} 2,5}}%
\put(2475,1100){\makebox(0,0){\strut{} 0}}%
\put(5415,1100){\makebox(0,0){\strut{} 0,2}}%
\put(8355,1100){\makebox(0,0){\strut{} 0,4}}%
\put(11295,1100){\makebox(0,0){\strut{} 0,6}}%
\put(14235,1100){\makebox(0,0){\strut{} 0,8}}%
\put(17175,1100){\makebox(0,0){\strut{} 1,0}}%
\put(550,5950){\rotatebox{90}{\makebox(0,0){\strut{}Current, $\bar I$}}}%
\put(9825,275){\makebox(0,0){\strut{}Reduced flux $\Phi/\Phi_0$}}%
\put(4700,9675){\makebox(0,0)[l]{\strut{}$\Gamma_0=0,01$}}%
\put(4700,9125){\makebox(0,0)[l]{\strut{}$\Gamma_0=0,1$}}%
\put(4700,8575){\makebox(0,0)[l]{\strut{}$\Gamma_0=0,5$}}%
\put(4700,8025){\makebox(0,0)[l]{\strut{}$\Gamma_0=1,0$}}%
\end{picture}%
\endgroup
\caption{Current as a function of the reduced flux $\Phi/\Phi_0$, for coupling values $\Gamma_0=Nv_0^2/\lambda$. The current is given in units of $cE_F/\Phi_0$. The parameters are $E_1=0,99$ e $E_2=1,01$, $d=10^{-3}$, $\ed=1,0$, all in units of $E_F$.}
\label{fig2}
\end{figure}

\begin{figure}
\begin{picture}(0,0)%
\includegraphics[scale=0.7]{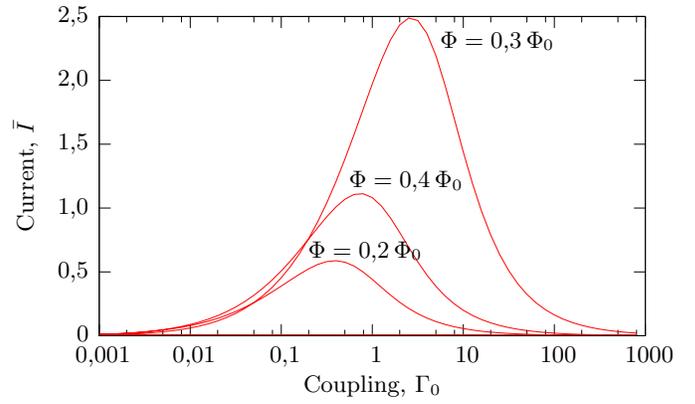}%
\end{picture}%
\begingroup
\setlength{\unitlength}{0.0140bp}%
\begin{picture}(18000,10800)(0,0)%
\put(2200,1650){\makebox(0,0)[r]{\strut{} 0}}%
\put(2200,3370){\makebox(0,0)[r]{\strut{} 0,5}}%
\put(2200,5090){\makebox(0,0)[r]{\strut{} 1,0}}%
\put(2200,6810){\makebox(0,0)[r]{\strut{} 1,5}}%
\put(2200,8530){\makebox(0,0)[r]{\strut{} 2,0}}%
\put(2200,10250){\makebox(0,0)[r]{\strut{} 2,5}}%
\put(2475,1100){\makebox(0,0){\strut{} 0,001}}%
\put(4925,1100){\makebox(0,0){\strut{} 0,01}}%
\put(7375,1100){\makebox(0,0){\strut{} 0,1}}%
\put(9825,1100){\makebox(0,0){\strut{} 1}}%
\put(12275,1100){\makebox(0,0){\strut{} 10}}%
\put(14725,1100){\makebox(0,0){\strut{} 100}}%
\put(17175,1100){\makebox(0,0){\strut{} 1000}}%
\put(550,5950){\rotatebox{90}{\makebox(0,0){\strut{}Current, $\bar I$}}}%
\put(9825,275){\makebox(0,0){\strut{}Coupling, $\Gamma_0$}}%
\put(8100,3900){\makebox(0,0)[l]{\strut{}$\Phi=0,\!2\,\Phi_0$}}%
\put(9200,5800){\makebox(0,0)[l]{\strut{}$\Phi=0,\!4\,\Phi_0$}}%
\put(11658,9562){\makebox(0,0)[l]{\strut{}$\Phi=0,\!3\,\Phi_0$}}%
\end{picture}%
\endgroup
\caption{Current as a function of the coupling $\Gamma_0$ in units of $cE_F/\Phi_0$. The parameters are $E_1=0,99$ e $E_2=1,01$, $d=10^{-3}$, $\ed=1,0$, all in units of $E_F$.}
\label{fig3}
\end{figure}

\begin{figure}
\begin{center}
\begin{picture}(0,0)%
\includegraphics[scale=0.7]{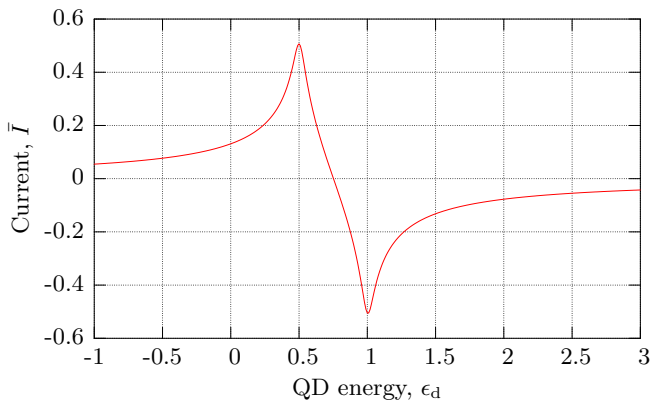}%
\end{picture}%
\begingroup
\setlength{\unitlength}{0.0140bp}%
\begin{picture}(18000,10800)(0,0)%
\put(2200,1650){\makebox(0,0)[r]{\strut{}-0.6}}%
\put(2200,3083){\makebox(0,0)[r]{\strut{}-0.4}}%
\put(2200,4517){\makebox(0,0)[r]{\strut{}-0.2}}%
\put(2200,5950){\makebox(0,0)[r]{\strut{} 0}}%
\put(2200,7383){\makebox(0,0)[r]{\strut{} 0.2}}%
\put(2200,8817){\makebox(0,0)[r]{\strut{} 0.4}}%
\put(2200,10250){\makebox(0,0)[r]{\strut{} 0.6}}%
\put(2475,1100){\makebox(0,0){\strut{}-1}}%
\put(4313,1100){\makebox(0,0){\strut{}-0.5}}%
\put(6150,1100){\makebox(0,0){\strut{} 0}}%
\put(7988,1100){\makebox(0,0){\strut{} 0.5}}%
\put(9825,1100){\makebox(0,0){\strut{} 1}}%
\put(11663,1100){\makebox(0,0){\strut{} 1.5}}%
\put(13500,1100){\makebox(0,0){\strut{} 2}}%
\put(15338,1100){\makebox(0,0){\strut{} 2.5}}%
\put(17175,1100){\makebox(0,0){\strut{} 3}}%
\put(550,5950){\rotatebox{90}{\makebox(0,0){\strut{}Current, $\bar I$}}}%
\put(9825,275){\makebox(0,0){\strut{}QD energy, $\ed$}}%
\end{picture}%
\endgroup
\caption{Current as a function of QD energy $\ed$ in units of $cE_F/\Phi_0$. The parameters are $E_1=0,99$ e $E_2=1,01$, $d=10^{-3}$,  $\Gamma_0=0,1$, all in units of $E_F$, and $\Phi=0,3\Phi_0$.}
\label{fig4}
\end{center}
\end{figure}

The plots of \fig{fig2} to \ref{fig4} show the current as function of different parameters as magnetic flux, resonance width $\Gamma_0$ and energy of the QD level. The energy of the QD can be tuned experimentally through the gate potential $V_g$. The natural energy unit adopted for the parameters is the Fermi energy of the ring $E_F$. For a semiconductor ring constituted of a GaAlAs/GaAs heterostructure of mesoscopic dimensions, i.e., with a diameter of $\mu m$, the total electron number is of thousands and the Fermi energy laying in the meV. The electron effective mass is about 0.07 the free electron mass\cite{lorke} and the elastic mean free path is about the circumference of the ring.

In the first plot we observe that the periodicity of the current is $\Phi_0$, double of the value found in the case without the QD. First, we should remember that the persistent current in a single isolated ring has a periodicity with the magnetic flux of  $\Phi_0$ both for the case of an odd or even number of electrons. What differs the current in each case is just a shift of $\Phi_0/2$. Thus, due to the very symmetry of those plots, when we perform an average over the number of electrons, i.e., we sum the currents in both cases, we end up with a current with a periodicity of $\Phi_0/2$. This is valid for a ring with or without disorder\cite{erlon-th}. The conclusion we can derive here is that the very presence of the QD breaks down the symmetry of the current for an even or odd number of electrons. Such asymmetry was also observed for a clean ring in the Kondo regime\cite{kang}, but now we can see that this feature is even more general.

Another remark to be done is about the coupling constant. For small values of this quantity ($\Gamma_0 \leq 10^{-3}E_F$), the current is basically a sine and it scales with $\Gamma_0$, such that $I/\Gamma_0$ is an universal function. For growing values of $\Gamma_0$ the current form begins to alter, the maximum (and minimum) of the curve approaches $\Phi_0/2$ and its amplitudes grows a bit slower with $\sqrt\Gamma_0$. In \fig{fig3} we see how the current behaves as a function of $\Gamma_0$ for some selected values of the magnetic flux. The first and obvious conclusion is that the current vanishes when $\Gamma_0\rightarrow0$, then this case is like we had opened the ring and no current is allowed to flow. In the same manner for $\Gamma_0$ very big, what we get again is a vanishing current. What happens now is something different. The life time of the resonance is inversely proportional to $\Gamma_0$ and therefore the probability of the electron to pass through the QD is diminished. If there is experimentally a controllable way to alter the coupling between the ring and the dot, then this plot shows where we can get a maximum amplitude for the current. 

Our main result, however, may be observed in \fig{fig4}, where we have analyzed the variation of the current amplitude as function of the energy of QD level. We have observed a peak in the current for $\ed$ near the Fermi energy of the ring. Such result is expected if we note that, in the case without the QD, the main contribution for the total current comes from the current associated with the electron in the uppermost level. Thus, when the energy of the QD coincides with this energy, there is a peak in the current indicating the resonance between both levels. The second peak however is quite curious. It occurs at about half of the Fermi energy and is symmetric to the first, only with opposite signal. The explanation for this phenomenon is the following. The coupling between the energy levels in the ring and that of the QD gives then a width and a Lorentzian shape. Since the current is essentially a derivative of the energy with respect to the magnetic flux, we get that the individual level contribution for the total current is proportional to the difference of its energy and $\ed$. Thus, when the QD level crosses down the Fermi energy, the contribution of the levels with higher energies than $\ed$ have opposite signal. For a given value of $\ed$, the total current vanish. Lowering $\ed$ even more, the current changes signal and increases until it reaches a maximum value and then decreases again as $\ed$ is much lower than the energy levels of the ring.

Unlikely the results obtained in other works\cite{xiong1,xiong2,kang,ferrari}, this extra peak is neither related to the Fano effect nor to the Kondo effect. The Fano effect\cite{fano,bulka,hofstetter} occurs when there is an interference between two paths through which the electron can pass, one with discrete levels, and the other a continuous band. To observe such effect, one may use one ring with a QD and two leads connected in opposite sides of the ring\cite{kobayashi}. Measuring the conductance we find peaks with asymmetric line shapes. On the other hand, Kondo effect\cite{hewson, ng, glazman} is related to the fact that the QD behaves like a magnetic impurity in the ring, and to observe it, it is necessary that the net spin in the dot be different of zero. As a result of the correlations between the electrons in the QD and those of the conduction band, there is a resonance near the Fermi energy with works as an extra channel for the electron tunneling. Our model does not account for the electronic spin and therefore there is no way of interpreting our result as a Kondo resonance, furthermore the second peak lays far below the Fermi level. In conclusion, the second peak arises as a constructive interference phenomenon between the levels of the ring and is purely a manifestation of the quantum nature of the system.

\begin{acknowledgments}
We thank C.H. Lewenkopf for helpful discussion. This work was supported by Brazilian agencies CNPq and CAPES.
\end{acknowledgments}


\begin{thebibliography}{18}
\expandafter\ifx\csname natexlab\endcsname\relax\def\natexlab#1{#1}\fi
\expandafter\ifx\csname bibnamefont\endcsname\relax
  \def\bibnamefont#1{#1}\fi
\expandafter\ifx\csname bibfnamefont\endcsname\relax
  \def\bibfnamefont#1{#1}\fi
\expandafter\ifx\csname citenamefont\endcsname\relax
  \def\citenamefont#1{#1}\fi
\expandafter\ifx\csname url\endcsname\relax
  \def\url#1{\texttt{#1}}\fi
\expandafter\ifx\csname urlprefix\endcsname\relax\def\urlprefix{URL }\fi
\providecommand{\bibinfo}[2]{#2}
\providecommand{\eprint}[2][]{\url{#2}}

\bibitem[{\citenamefont{Alhassid}(2000)}]{alhassid}
\bibinfo{author}{\bibfnamefont{Y.}~\bibnamefont{Alhassid}},
  \bibinfo{journal}{Rev. Mod. Phys.} \textbf{\bibinfo{volume}{72}},
  \bibinfo{pages}{895} (\bibinfo{year}{2000}).

\bibitem[{\citenamefont{Xiong and Liang}(2005)}]{xiong1}
\bibinfo{author}{\bibfnamefont{Y.-J.} \bibnamefont{Xiong}} \bibnamefont{and}
  \bibinfo{author}{\bibfnamefont{X.-T.} \bibnamefont{Liang}},
  \bibinfo{journal}{Physica B} \textbf{\bibinfo{volume}{355}},
  \bibinfo{pages}{216} (\bibinfo{year}{2005}).

\bibitem[{\citenamefont{Xiong and Liang}(2004)}]{xiong2}
\bibinfo{author}{\bibfnamefont{Y.-J.} \bibnamefont{Xiong}} \bibnamefont{and}
  \bibinfo{author}{\bibfnamefont{X.-T.} \bibnamefont{Liang}},
  \bibinfo{journal}{Phys. Lett. A} \textbf{\bibinfo{volume}{330}},
  \bibinfo{pages}{307} (\bibinfo{year}{2004}).

\bibitem[{\citenamefont{Kang and Shin}(2000)}]{kang}
\bibinfo{author}{\bibfnamefont{K.}~\bibnamefont{Kang}} \bibnamefont{and}
  \bibinfo{author}{\bibfnamefont{S.-C.} \bibnamefont{Shin}},
  \bibinfo{journal}{Phys. Rev. Lett.} \textbf{\bibinfo{volume}{85}},
  \bibinfo{pages}{5619} (\bibinfo{year}{2000}).

\bibitem[{\citenamefont{Ferrari et~al.}(1999)\citenamefont{Ferrari, Chiappe,
  Anda, and Davidovich}}]{ferrari}
\bibinfo{author}{\bibfnamefont{V.}~\bibnamefont{Ferrari}},
  \bibinfo{author}{\bibfnamefont{G.}~\bibnamefont{Chiappe}},
  \bibinfo{author}{\bibfnamefont{E.}~\bibnamefont{Anda}}, \bibnamefont{and}
  \bibinfo{author}{\bibfnamefont{M.~A.} \bibnamefont{Davidovich}},
  \bibinfo{journal}{Phys. Rev. Lett.} \textbf{\bibinfo{volume}{82}},
  \bibinfo{pages}{5088} (\bibinfo{year}{1999}).

\bibitem[{\citenamefont{Kobayashi et~al.}(2004)\citenamefont{Kobayashi, Aikawa,
  Katsumoto, and Iye}}]{kobayashi}
\bibinfo{author}{\bibfnamefont{K.}~\bibnamefont{Kobayashi}},
  \bibinfo{author}{\bibfnamefont{H.}~\bibnamefont{Aikawa}},
  \bibinfo{author}{\bibfnamefont{S.}~\bibnamefont{Katsumoto}},
  \bibnamefont{and} \bibinfo{author}{\bibfnamefont{Y.}~\bibnamefont{Iye}},
  \bibinfo{journal}{Physica E} \textbf{\bibinfo{volume}{22}},
  \bibinfo{pages}{468} (\bibinfo{year}{2004}).

\bibitem[{\citenamefont{Hewson}(1993)}]{hewson}
\bibinfo{author}{\bibfnamefont{A.~C.} \bibnamefont{Hewson}},
  \emph{\bibinfo{title}{The Kondo Problem to Heavy Fermions}}
  (\bibinfo{publisher}{Cambridge University Press},
  \bibinfo{address}{Cambridge}, \bibinfo{year}{1993}).

\bibitem[{\citenamefont{Ng et~al.}(1988)}]{ng}
\bibinfo{author}{\bibfnamefont{T.}~\bibnamefont{Ng}} \bibnamefont{et~al.},
  \bibinfo{journal}{Phys. Rev. Lett.} \textbf{\bibinfo{volume}{61}},
  \bibinfo{pages}{1768} (\bibinfo{year}{1988}).

\bibitem[{\citenamefont{Glazman et~al.}(1988)}]{glazman}
\bibinfo{author}{\bibfnamefont{L.~I.} \bibnamefont{Glazman}}
  \bibnamefont{et~al.}, \bibinfo{journal}{JETP Lett.}
  \textbf{\bibinfo{volume}{47}}, \bibinfo{pages}{452} (\bibinfo{year}{1988}).

\bibitem[{\citenamefont{Iida et~al.}(1990)\citenamefont{Iida, Weidenm{\"u}ller,
  and Zuk}}]{iwz}
\bibinfo{author}{\bibfnamefont{S.}~\bibnamefont{Iida}},
  \bibinfo{author}{\bibfnamefont{H.~A.} \bibnamefont{Weidenm{\"u}ller}},
  \bibnamefont{and} \bibinfo{author}{\bibfnamefont{J.~A.} \bibnamefont{Zuk}},
  \bibinfo{journal}{Ann. Phys. (NY)} \textbf{\bibinfo{volume}{200}},
  \bibinfo{pages}{219} (\bibinfo{year}{1990}).

\bibitem[{\citenamefont{Altland et~al.}(1992)\citenamefont{Altland, Iida,
  M{\"u}ller-Groeling, and Weidenm{\"u}ller}}]{altland}
\bibinfo{author}{\bibfnamefont{A.}~\bibnamefont{Altland}},
  \bibinfo{author}{\bibfnamefont{S.}~\bibnamefont{Iida}},
  \bibinfo{author}{\bibfnamefont{A.}~\bibnamefont{M{\"u}ller-Groeling}},
  \bibnamefont{and} \bibinfo{author}{\bibfnamefont{H.~A.}
  \bibnamefont{Weidenm{\"u}ller}}, \bibinfo{journal}{Ann. Phys. (NY)}
  \textbf{\bibinfo{volume}{219}}, \bibinfo{pages}{148} (\bibinfo{year}{1992}).

\bibitem[{\citenamefont{Verbaarschot et~al.}(1985)\citenamefont{Verbaarschot,
  Weidenm{\"u}ller, and Zirnbauer}}]{vwz}
\bibinfo{author}{\bibfnamefont{J.~J.~M.} \bibnamefont{Verbaarschot}},
  \bibinfo{author}{\bibfnamefont{H.~A.} \bibnamefont{Weidenm{\"u}ller}},
  \bibnamefont{and}
  \bibinfo{author}{\bibfnamefont{M.}~\bibnamefont{Zirnbauer}},
  \bibinfo{journal}{Phys. Rep.} \textbf{\bibinfo{volume}{129}},
  \bibinfo{pages}{367} (\bibinfo{year}{1985}).

\bibitem[{\citenamefont{Altland et~al.}(1993)\citenamefont{Altland, Iida, and
  Efetov}}]{altland2}
\bibinfo{author}{\bibfnamefont{A.}~\bibnamefont{Altland}},
  \bibinfo{author}{\bibfnamefont{S.}~\bibnamefont{Iida}}, \bibnamefont{and}
  \bibinfo{author}{\bibfnamefont{K.~B.} \bibnamefont{Efetov}},
  \bibinfo{journal}{J. Phys. A} \textbf{\bibinfo{volume}{26}},
  \bibinfo{pages}{3545} (\bibinfo{year}{1993}).

\bibitem[{\citenamefont{Lorke et~al.}(2000)}]{lorke}
\bibinfo{author}{\bibfnamefont{A.}~\bibnamefont{Lorke}} \bibnamefont{et~al.},
  \bibinfo{journal}{Phys. Rev. Lett.} \textbf{\bibinfo{volume}{84}},
  \bibinfo{pages}{2223} (\bibinfo{year}{2000}).

\bibitem[{\citenamefont{Ferreira}(2004)}]{erlon-th}
\bibinfo{author}{\bibfnamefont{E.~H.~M.} \bibnamefont{Ferreira}}, Ph.D. thesis,
  \bibinfo{school}{Dep. Fisica, Universidade Federal de Minas Gerais}
  (\bibinfo{year}{2004}).

\bibitem[{\citenamefont{Fano}(1961)}]{fano}
\bibinfo{author}{\bibfnamefont{U.}~\bibnamefont{Fano}}, \bibinfo{journal}{Phys.
  Rev.} \textbf{\bibinfo{volume}{124}}, \bibinfo{pages}{1866}
  (\bibinfo{year}{1961}).

\bibitem[{\citenamefont{Bulka et~al.}(2001)}]{bulka}
\bibinfo{author}{\bibfnamefont{B.}~\bibnamefont{Bulka}} \bibnamefont{et~al.},
  \bibinfo{journal}{Phys. Rev. Lett.} \textbf{\bibinfo{volume}{86}},
  \bibinfo{pages}{5128} (\bibinfo{year}{2001}).

\bibitem[{\citenamefont{Hofstetter et~al.}(2001)}]{hofstetter}
\bibinfo{author}{\bibfnamefont{W.}~\bibnamefont{Hofstetter}}
  \bibnamefont{et~al.}, \bibinfo{journal}{Phys. Rev. Lett.}
  \textbf{\bibinfo{volume}{87}}, \bibinfo{pages}{156803}
  (\bibinfo{year}{2001}).

\end{thebibliography}

\end{document}